\documentstyle[aps,epsbox]{revtex}
\begin{document}
\title{On integrability test for ultradiscrete equations}
\author{Daisuke Takahashi\dag and Kenji Kajiwara\ddag}
\address{\dag Department of Mathematical Sciences, Waseda University,
Ohkubo 3-4-1, Shinjuku-ku, Tokyo 169-8555, JAPAN\\ daisuke@mse.waseda.ac.jp}
\address{\ddag Department of Electrical Engineering, Doshisha University,
Kyotanabe, Kyoto 610-0321, JAPAN \\
kaji@elrond.doshisha.ac.jp}
\date{\today}
\maketitle
\begin{abstract} 
  We consider an integrability test for ultradiscrete equations based on the
singularity confinement analysis for discrete equations.  We show how
singularity pattern of the test is transformed into that of ultradiscrete
equation.  The ultradiscrete solution pattern can be interpreted as a
perturbed solution.  We can also check an integrability of a given equation
by a perturbation growth of a solution, namely Lyapunov exponent.  Therefore,
singularity confinement test and Lyapunov exponent are related each other in
ultradiscrete equations and we propose an integrability test from this point
of view.
\end{abstract}
\section{Introduction}
Integrability is an important concept in nonlinear equations.  If a given
equation turns out to be integrable, we can get many exact structures from
the system, for example, conserved quantities, symmetries, exact solutions,
and so on.  Therefore, it has been an important problem to test integrability
of equations.  For differential equations, it is well known that the
Painlev\'e test is powerful to detect integrability.~\cite{Painleve} However,
for difference equations, the Painlev\'e test cannot be applied due to
discreteness of independent variables, and the singularity confinement (SC)
test is proposed instead.~\cite{SC}

Let us consider the following multiplicative type of difference equation
\begin{equation}  \label{dp}
    x_{n+1} x_n^\sigma x_{n-1} = \alpha \lambda^n x_n + 1.
\end{equation}
This equation for $\sigma=0,1,2$ with $\lambda=1$ belongs to so-called 
Quispel-Roberts-Thompson (QRT) system~\cite{QRT1,QRT2} which is a
large family of integrable second order ordinary difference equations. The
term ``integrability'' is somewhat more delicate in discrete system than
continuous one. For QRT system, it is integrable in a sense that it has a
quartic conserved quantity and thus the general solution is expressed by
elliptic functions. For generic $\lambda$, equation (\ref{dp}) with $\sigma=0$,
1, 2 are known as discrete analogues to Painlev\'e I equation, which are
considered to be integrable in a sense that it passes the SC
test.~\cite{ultra painleve}  It is considered to be non-integrable for other
$\sigma$. 

The SC test is applied to equation (\ref{dp}) as follows.  Let
us assume $\sigma=2$ and $\lambda=1$ to make discussions easier.
If initial $x$'s(assume $x_0$ and $x_1$) are $x_0=f$ (non-zero finite) and
$x_1=-\frac{1}{\alpha}$ respectively, then we have $x_2=0$, and $x_3$ becomes
singular. In order to see the behaviour of this singularity, we introduce a small parameter
$\delta$ ($\sim 0$) and put $x_0=f$ and $x_1=-\frac{1}{\alpha}+\delta$. 
Then, successive iteration and  asymptotic evaluation in $\delta$ gives a singularity
pattern as follows:
\begin{equation}  \label{SCsig2}
\begin{tabular}{|c|c|c|c|c|c|c|c|}
\hline
  $x_0$ & $x_1$ & $x_2$ & $x_3$ & $x_4$ & $x_5$ & $x_6$ & $x_7$ \\
\hline
  $f$ &
  $-\frac{1}{\alpha}+\delta$ &
  $\frac{\alpha^3}{f}\delta$ &
  $-\frac{f^2}{\alpha^5} \delta^{-2}$ &
  $-\frac{\alpha^3}{f}\delta$ &
  $-\frac{1}{\alpha}-\delta$ &
  $f$ &
  $-\frac{\alpha+\alpha^2f}{f^2}$ \\
\hline
\end{tabular}
\end{equation}
The above pattern shows;(a) A singularity due to $x_1$ occurs at $n=3$,
(b) The singularity is confined, that is, does not spread on the whole
lattice, (c) Information on $x_0$ pass through the singularity to
$x_n$ at $n\geq4$.  Following to SC test, both locality of singularity
and preservation of information on initial data strongly indicate an
integrability of the equation.

There is another singularity pattern using initial data $x_0=f$ and $x_1=\delta$:
\begin{equation}  \label{SCsig2-2}
\hspace*{-60pt}
\begin{tabular}{|c|c|c|c|c|c|c|c|c|c|}
\hline
  $x_0$ & $x_1$ & $x_2$ & $x_3$ & $x_4$ & $x_5$ & $x_6$ & $x_7$ & $x_8$ & $x_9$ \\
\hline
  $f$ &
  $\delta$ &
  $\frac{1}{f}\delta^{-2}$ &
  $\alpha\,f\,\delta$ &
  $\frac{1}{\alpha^2\,f}$ &
  $\alpha^2(1+\alpha\,f)\delta^{-1}$ &
  $\frac{\alpha\,f}{1+\alpha\,f}\delta$ &
  $\frac{1+\alpha\,f}{\alpha^4f^2}\delta^{-1}$ &
  $\alpha^4f$ &
  $\frac{1+\alpha^5f}{\alpha^4(1+\alpha\,f)}\delta$ \\
\hline
\end{tabular}
\end{equation}
The above SC pattern shows; (a) $x_{8n}=\text{finite}$ and
$x_{8n+1}=O(\delta)$ for any $n$, (b) Information of $x_0$ pass
through periodic singularities to $x_n$ at large $n$.  In this
pattern, singularity spreads on the whole lattice but singular points
are confined by finite values which have information on initial $x_0$.
Although such periodic singularity pattern
has not been discussed anywhere to the authors' knowledge, we may consider that this pattern
also suggests integrability.

Next let us consider an ultradiscrete analogue to equation
(\ref{dp}).~\cite{Box and Ball,ultra Toda,ultra Toda 2,u-LV,ultra painleve,ultra elliptic,ultra mkdv,ultra
chaos,ultra Burgers,ultra painleve 2}  Applying the following
transformation,
\begin{equation}  \label{trans}
  x_n = e^{X_n / \varepsilon}, \qquad \alpha = e^{A/\varepsilon}, \qquad \lambda=e^{L/\varepsilon},
\end{equation}
we obtain
\begin{equation}
  X_{n+1} + \sigma\,X_n + X_{n-1} = \varepsilon \log(1 + e^{(X_n+n\,L+A)/\varepsilon}).
\end{equation}
If we take a limit $\varepsilon\to+0$, we get
\begin{equation}  \label{up}
  X_{n+1} + \sigma\,X_n + X_{n-1} = \max(0,\,X_n+n\,L+A),
\end{equation}
using a formula
\begin{equation}  \label{ultra limit}
  \lim_{\varepsilon\to+0} \varepsilon\,\log(e^{a/\varepsilon} + e^{b/\varepsilon} + \cdots) = \max(a,\,b,\,\cdots).
\end{equation}
If initial data $X_0$, $X_1$ and parameters $L$ and $A$ of (\ref{up})
are all integer, $X_n$ for any $n$ is always integer.  In this sense,
the dependent variable of equation (\ref{up}) is discretized
through a limit (\ref{ultra limit}).  The above discretization process
on dependent variable is called 'ultradiscretization' and equation (\ref{up})
is an ultradiscrete analogue to equation. (\ref{dp})~\cite{u-LV}
We obtain a cellular automaton from an ultradiscrete equation, if it is possible
to restrict values of dependent variable
in a range of integers. Thus, it is expected that
ultradiscrete equations connect the continuous and the discrete worlds,
that is, differential or difference equations and cellular automata.

A large difference between ultradiscrete equations and discrete ones is an
existence of singularities.  Since ultradiscrete equations are
piecewise linear, singularities in normal sense do not exist.  Therefore, we cannot
use SC test as it is. 

In this letter, we demonstrate our trial to detect integrability of
ultradiscrete equations based on the SC test for discrete equations.  In
the next section, we show how SC test is transformed in equation
(\ref{up}).  Then, we discuss differences between ultradiscrete
analogous to integrable difference equations and those to nonintegrable
ones.  In section 3, we show that SC test is related to Lyapunov
exponent in ultradiscrete equations.  Lyapunov exponent is a growth rate
of perturbation of solution, by which we can test integrability of a
given equation.~\cite{chaos} In the last section, we give concluding
remarks.
\section{Ultradiscrete analogue to singularity pattern}
In this section, we discuss how the singularity pattern of the difference
equation (\ref{dp}) is transformed to that of ultradiscrete analogue (\ref{up}).
First we note that the case of $\sigma=0,1,2$ in autonomous case ($L=0$) of
equation (\ref{up}) should be considered to be integrable, since;(a) Each of
them has a conserved quantity (b) Each of them admits a general solution which
is expressed by ultradiscrete analogue of elliptic
functions.~\cite{ultra elliptic} Therefore, we can examine whether a
transformed singularity pattern is useful to test an integrability of
ultradiscrete equation or not by comparing the pattern for $\sigma=0,1,2$ with
that for other cases.

When we ultradiscretize equation (\ref{dp}) to obtain equation (\ref{up}), we
use transformation of variables (\ref{trans}).  In this transformation, if
$L$ and $A$ are not 0, this means that we introduce singularities into
parameters $\alpha$ and $\lambda$ as well as $x$ in equation (\ref{dp}).
In order to discuss effects of singular parameters separately, first we
consider a case of finite $\alpha$ and $\lambda$, that is, $A=L=0$.  Moreover,
let us assume $\sigma=2$ case,
\begin{equation}  \label{upsig2a0l0}
    X_{n+1} + 2\,X_n + X_{n-1} = \max(0,\,X_n),
\end{equation}
as a concrete example.  In the transformation (\ref{trans}), $x$ (also
$\alpha$ and $\lambda$) must be always positive.  Therefore, we cannot 
obtain an ultradiscrete pattern corresponding to pattern (\ref{SCsig2})
because $x_n$ necessarily becomes negative at certain $n$'s.

On the other hand, if $f$ and $\delta$ are positive, $x_n$ in the
singularity pattern (\ref{SCsig2-2}) is always positive and can be
ultradiscretized.  In the limiting process, relation between those
values and ultradiscretization parameter $\varepsilon$ is
important.  Let us assume that $f$ is finite independent of $\varepsilon$
and $\delta=e^{-K/\varepsilon}$ where $K$ is positive and finite.
Then, we get the following pattern on $X_n$ from pattern (\ref{SCsig2-2})
through (\ref{trans}).
\begin{equation}  \label{uSCsig2}
\begin{tabular}{|c|c|c|c|c|c|c|c|c|c|}
\hline
  $X_0$ & $X_1$ & $X_2$ & $X_3$ & $X_4$ & $X_5$ & $X_6$ & $X_7$ & $X_8$ & $X_9$ \\
\hline
  $0$ & $-K$ & $2K$ & $-K$ & $0$ & $K$ & $-K$ & $K$ & $0$ & $-K$ \\
\hline
\end{tabular}
\end{equation}
This pattern is strictly periodic with period 8.  Note that this pattern
can also be obtained from equation (\ref{upsig2a0l0}) with initial data $X_0=0$
and $X_1=-K$ directly.  We can easily see that positive, zero,
negative $X_n$ correspond to infinite, finite, infinitesimal $x_n$
respectively.  Periodic confinement of singularities of $x_n$ is
transformed into separation of positive values by 0's in $X_n$.
However, only $K$ corresponding to $\delta$ survives and information
on $f$ is lost in the pattern.

It is important in the SC test whether the information of initial data survives
or not. Therefore, next we take $f=e^{\rho/\varepsilon}$ and
$\delta=e^{-K/\varepsilon}$ where $\rho$ is finite and $|\rho|\ll K$.  The
initial data become $X_0=\rho$ and $X_1=-K$, and the following pattern
is obtained from pattern (\ref{SCsig2-2}), 
\begin{equation}  \label{uSCsig2-2}
\begin{array}{l}
\begin{tabular}{|c|c|c|c|c|c|}
\hline
  $X_0$ & $X_1$ & $X_2$ & $X_3$ & $X_4$ & $X_5$ \\
\hline
  $\rho$ & $-K$ & $2K-\rho$ & $-K+\rho$ & $-\rho$
  & $K+\max(0,\rho)$\\
\hline
\end{tabular}\\
\begin{tabular}[t]{|c|c|c|c|}
\hline
 $X_6$ & $X_7$ & $X_8$ & $X_9$\\
\hline
   $\begin{array}{c} -K+\rho \\ -\max(0,\rho)\end{array}$
  & $\begin{array}{c} K-2\rho \\ +\max(0,\rho)\end{array}$
  & $\rho$  & $-K$ \\
\hline
\end{tabular}
\end{array}
\end{equation}
This pattern is also periodic with period 8.  The information of initial
data, namely $\rho$, clearly survives for large $n$.  It is easy to see that 
we can obtain this pattern directly by successive iteration of
equation~(\ref{upsig2a0l0}).  Therefore, we can conclude that the equation and
its singular solution pattern are consistently transformed from difference
equation to ultradiscrete one at least in this example.  Similar results are
obtained for the other integrable cases ($\sigma=0$ and 1).

Next, we test non-integrable case of equation (\ref{dp}), $\sigma=3$.
Let us assume $\lambda=1$ for simplicity, then equation (\ref{dp}) becomes
\begin{equation}  \label{dpsig3l1}
    x_{n+1} x_n^3 x_{n-1} = \alpha x_n + 1,
\end{equation}
and we get a singularity pattern
\begin{equation}  \label{SCsig3}
\hspace*{-40pt}
\begin{tabular}{|c|c|c|c|c|c|c|c|}
\hline
  $x_0$ & $x_1$ & $x_2$ & $x_3$ & $x_4$ & $x_5$ & $x_6$ & $x_7$ \\
\hline
  $f$ &
  $\delta$ &
  $\frac{1}{f}\,\delta^{-3}$ &
  $\alpha\,f^2\delta^5$ &
  $\frac{1}{\alpha^3f^5}\,\delta^{-12}$ &
  $\alpha^6f^8\delta^{19}$ &
  $\frac{1}{\alpha^{15}f^{19}}\,\delta^{-45}$ &
  $\alpha^{25}f^{30}\delta^{71}$ \\
\hline
\end{tabular}
\end{equation}
with $x_0=f$ and $x_1=\delta$.  We can easily prove that
$x_n=O(\delta^{p_n})$ and
$p_{2n}\sim-\frac{\sqrt{3}}{2}((2+\sqrt{3})^n-(2-\sqrt{3})^n)$ and
$p_{2n+1}\sim\frac{\sqrt{3}+1}{2}(2+\sqrt{3})^n-\frac{\sqrt{3}-1}{2}(2-\sqrt{3})^n$.
Singularities are not confined and information on initial data is lost
for large $n$ when $\delta\to0$.  Following to SC test, this strongly implies
that equation (\ref{dpsig3l1}) is not integrable.

The corresponding ultradiscrete equation is
\begin{equation}  \label{upsig3a0l0}
    X_{n+1} + 3\,X_n + X_{n-1} = \max(0,\,X_n).
\end{equation}
and a pattern becomes
\begin{equation}  \label{SCsig3a0l0}
\begin{array}{l}
\begin{tabular}{|c|c|c|c|c|c|}
\hline
  $X_0$ & $X_1$ & $X_2$ & $X_3$ & $X_4$ & $X_5$ \\
\hline
  $\rho$ & $-K$ & $3K-\rho$ & $-5K+2\rho$
  & $12K-5\rho$ & $-19K+8\rho$ \\
\hline
\end{tabular}  \\
\begin{tabular}[t]{|c|c|c|c|}
\hline
  $X_6$ & $X_7$ & $X_8$ & $X_9$ \\
\hline
  $45K-19\rho$ & $71K+30\rho$ & $-187K-41\rho$ & $490K+93\rho$ \\
\hline
\end{tabular}
\end{array}
\end{equation}
In this pattern, the first term including $K$ in every $X_n$ ($n\geq1$)
becomes dominant and information on $\rho$ becomes negligible.

In the above examples, we have shown singularity pattern of difference
equation (\ref{dp}) for finite $\alpha$ with $\lambda=1$ and
corresponding pattern of ultradiscrete equation (\ref{up})
simultaneously.  The values $f$ (finite) and $1/\delta$ (singular) in
equation~(\ref{dp}) corresponds to $\rho$ and $K$ ($|\rho|\ll K$) in
equation~(\ref{up}), respectively.  When equation (\ref{dp}) is integrable
($\sigma=0, 1, 2$), information of $\rho$ survives for large $n$ in
corresponding ultradiscrete equations.  Confined singularity in the SC test on
difference equation (\ref{dp}) corresponds to separation of positive values
including $K$ by small $\rho$.

On the other hand, when equation (\ref{dp})
is non-integrable ($\sigma=3$), information on $\rho$ is negligible
and $K$ becomes dominant for large $n$ in corresponding ultradiscrete equation,
and separation of positive values by $\rho$ does not occur.

We note that the above singular solution patterns for ultradiscrete equation
are obtained without inconsistency. Namely, the pattern obtained by taking
ultradiscrete limit on the singularity pattern of difference equation always
coincides with that obtained by successive iteration of ultradiscrete equation
directly.

Therefore, we could obtain an interpretation of the SC test on the
ultradisctete equation (\ref{up}) with $L=A=0$, and observe a critical
difference between the case of $\sigma=0,1,2$ and other cases.
\section{Lyapunov exponent of perturbed ultradiscrete solution}
\begin{figure}
\begin{center}
  \epsfile{file=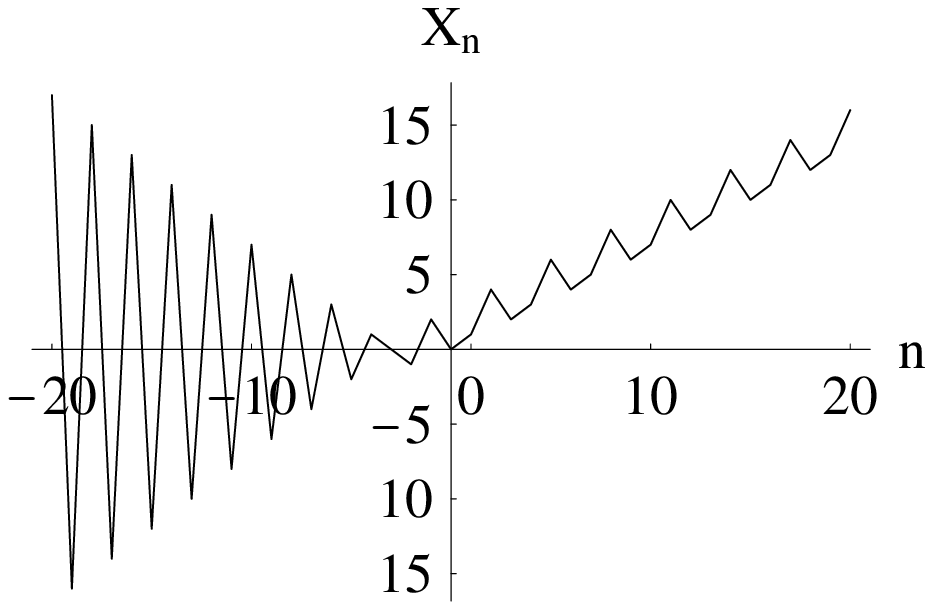}
\end{center}
  \caption{Plot of $X_n$ to (\ref{upsig2}) for $A=3$, $L=2$, $X_0=0$ and $X_1=1$.}
  \label{fig1}
\end{figure}

So far, we have considered the ultradiscrete equation (\ref{up}) with $L=A=0$.
However, the parameters $L$ and $A$ can effect a solution drastically. 
For example, let us take $\sigma=2$ case in equation (\ref{up}) again,
\begin{equation}  \label{upsig2}
    X_{n+1} + 2\,X_n + X_{n-1} = \max(0,\,X_n+n\,L+A),
\end{equation}
and put $L=0$ and $A=1$. Let us start with the initial data $X_0=0$ and
$X_1=-1$. Then we see that this solution is periodic with period 3. However,
starting with $X_0=0$ and $X_1=-2$, we see that the period is now 20. Compare
this phenomenon with a fact that the period for $L=A=0$ was 8 for any
$X_0=\rho$ and $X_1=-K$. Therefore, the period of solution can change by the
combination of parameter $A$ and initial data $X_0$, $X_1$. 
Moreover, in a case of $L\ne 0$, equation (\ref{upsig2}) becomes
non-autonomous equation. Therefore, since the values of $X_n$'s grow due to
the term $nL$, both criterions obtained in the previous section do not work
as they are. That is, information on initial value becomes negligible and
separation of positive values by $\rho$ does not occur. A solution to equation
(\ref{upsig2}) in the case of $A=3$, $L=2$, $X_0=0$, $X_1=1$ is shown in
Fig.~\ref{fig1}.

These observations imply that we cannot simply transform a singularity
pattern of difference equation to ultradiscrete one for $L\neq 0$ and
$A\neq 0$.  This is because taking the parameters $L$ and $A$ to be
non-zero means that we have introduced singularities on parameter
$\lambda$ and $\alpha$ in the difference equation (\ref{dp}),
respectively. Therefore, we must improve an integrability test for
ultradiscrete equation.

Let us look back the singular solution patterns (\ref{uSCsig2-2}) and
(\ref{SCsig3a0l0}) of the ultradiscrete equations (\ref{upsig2a0l0}) and
(\ref{upsig3a0l0}), respectively.
We can give another interpretation of $\rho$ and $K$.  $K$ is a parameter
giving a solution orbit with initial data $X_0=0$, $X_1=-K$ and $\rho$ is
a perturbation to the orbit.  From this viewpoint,
growth rates of perturbation in the patterns (\ref{uSCsig2-2}) and
(\ref{SCsig3a0l0}) are clearly different.
Therefore, it might be possible to regard SC analysis on ultradiscrete
equations is nothing but analysis on a growth rate of perturbation.

\begin{figure}
\begin{center}
  \epsfile{file=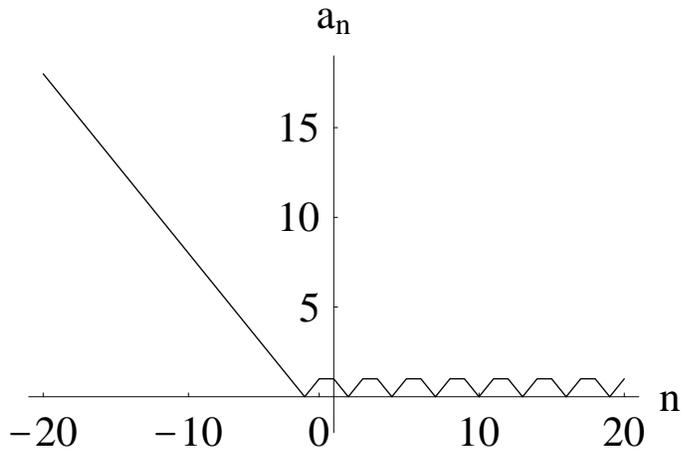}
\end{center}
  \caption{Plot of $a_n$ to (\ref{upsig2}) for $A=3$, $L=2$, $X_0=0$, $X'_0=\rho$, and $X_1=X'_1=1$.}
  \label{fig2}
\end{figure}

Let us assume that $X_n$ is a solution to equation (\ref{upsig2})
and $X'_n$ is one perturbed by $\rho$.
Then, amplitude of perturbation is defined by $a_n$
where $a_n\equiv|\frac{X_n-X'_n}{\rho}|$.  Figure~\ref{fig2} shows
$a_n$ for $A=3$, $L=2$, $X_0=0$, $X'_0=\rho$ and $X_1=X'_1=1$.  We can
see that $a_n$ grows linearly for $n<0$ and does not grow for $n>0$.
Therefore, the solution is not chaotic and we can easily estimate a
global behavior of the solution.

\begin{figure}
\begin{center}
  \epsfile{file=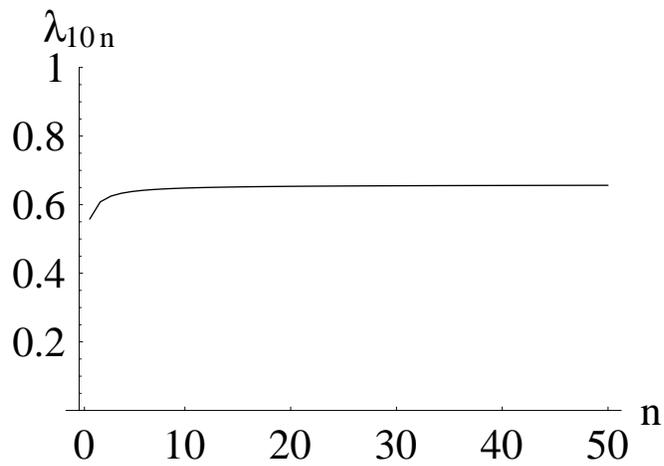}
\end{center}
  \caption{Plot of $\lambda_{10n}$ to (\ref{upsig3a0l0}) for $X_0=0$, $X'_0=\rho$, and $X_1=X'_1=1$.}
  \label{fig3}
\end{figure}

A quantitative index to check integrability of a system is Lyapunov exponent
which is a mean growth rate of perturbation.~\cite{chaos}  If we define
$\lambda_n$ by $\frac{1}{n}\log a_n$, $\lambda_\infty$
($\lambda_{-\infty}$) is the exponent.  When $\lambda_\infty>0$, the
perturbation grows exponentially and the system becomes chaotic.  As
for equation (\ref{upsig3a0l0}), $\lambda_{10n}$ for $X_0=0$, $X'_0=\rho$ and
$X_1=X'_1=1$ is plotted in Fig.~\ref{fig3}.  $\lambda_n$ is
asymptotically about 0.66 for large $n$.  Therefore, we can conclude
that equation (\ref{upsig3a0l0}) is chaotic and this result is compatible with
that of the previous section.

Next we try another example, the cutoff Toda equation~\cite{cutoff Toda}
\begin{equation}  \label{cutoff Toda}
\left\{\begin{array}{rl}
  x_{n+1} &= x_n/y_n^\sigma \\
  y_{n+1} &= y_n\,(h+x_{n+1})/(h+1/x_{n+1})
\end{array}\right.,
\end{equation}
where $\sigma$ and $h$ are constant parameters.  Eliminating $y$, this equation
reduces to the following ordinary difference equation on $x_n$,
\begin{equation}  \label{dp2}
  x_{n+1}x_{n-1} = x_n^{2-\sigma}(1+h\,x_n)^\sigma/(h+x_n)^\sigma.
\end{equation}
Equation (\ref{dp2}) with $\sigma=1$ and that with $\sigma=2$ are nothing but
the autonomous
discrete Painlev\'e II  and III equations, respectively.~\cite{ultra painleve}  According
to SC test, only the cases of $\sigma=1$ and 2 are integrable and not
otherwise.  Here we discuss details of perturbed solutions of
ultradiscrete analogue to (\ref{cutoff Toda}) in the cases of $\sigma=2$
(integrable) and 3 (non-integrable).

If we introduce $x_n=\exp(X_n/\varepsilon)$, $h=\exp(H/\varepsilon)$ and take
the limit $\varepsilon\to+0$, we obtain
\begin{equation}  \label{ultra cutoff}
  X_{n+1} + X_{n-1} = (2-\sigma)\,X_n + \sigma\,(\max(0,\,X_n+H)-\max(H,\,X_n)),
\end{equation}
from equation (\ref{dp2}).  If $H\ne0$, this means that equation (\ref{dp2})
includes singular parameter
$h$.  In the case of $\sigma=2$, we see that the difference
equation~(\ref{dp2}) has the following conserved quantity,
\begin{equation}
  x_{n+1}x_n + 2h(x_{n+1}+x_n) +
  h^2(\frac{x_{n+1}}{x_n}+\frac{x_n}{x_{n+1}}) +
  2h(\frac{1}{x_{n+1}}+\frac{1}{x_n}) + \frac{1}{x_{n+1}x_n}.
\end{equation}
Then, we obtain a conserved quantity of ultradiscrete
equation~(\ref{ultra cutoff}) with $\sigma=2$,
\begin{equation}
  \max(X_n+X_{n+1},\,-X_n-X_{n+1},\,2H-X_n+X_{n+1},\,2H+X_n-X_{n+1}),
\end{equation}
through the above ultradiscrete limit.  Define $K$ by
$\max(X_0+X_1,\,-X_0-X_1,\,2H-X_0+X_1,\,2H+X_0-X_1)$, then any point
$(X_n,\,X_{n+1})$ ($n\geq0$) in phase space always exists on sides of
a rectangle with vertices $(H-K,\,-H)$, $(H,\,K-H)$, $(-H,\,H-K)$,
$(K-H,\,H)$.  If we use perturbed initial data $(X_0+\rho_1,\,
X_1+\rho_2)$ where $\rho_1\sim0$ and $\rho_2\sim0$, then an orbit of
$(X_n,\,X_{n+1})$ shifts infinitesimally but it is still stable.

To estimate a growth of perturbation, let us calculate amplitude of
perturbation $a_n$ by solutions $X_n$ and $X'_n$ which are the solutions from
given initial data $(X_0,\,X_1)$ and $(X_0+\rho,\,X_1)$, respectively.
If we choose $H=5$, $X_0=0$ and $X_1=1$, then $X_n$ becomes periodic
with period 24 and $a_{24n}=20n-1$.  If we take $H=10$, $X_0=11$ and
$X_1=17$, period is 9 and $a_{9n}=5n+1$.  In both cases, $a_n$ grows
linearly on $n$.  These observations suggest that equation (\ref{ultra cutoff})
with $\sigma=2$ is integrable.

\begin{figure}
\begin{center}
  \epsfile{file=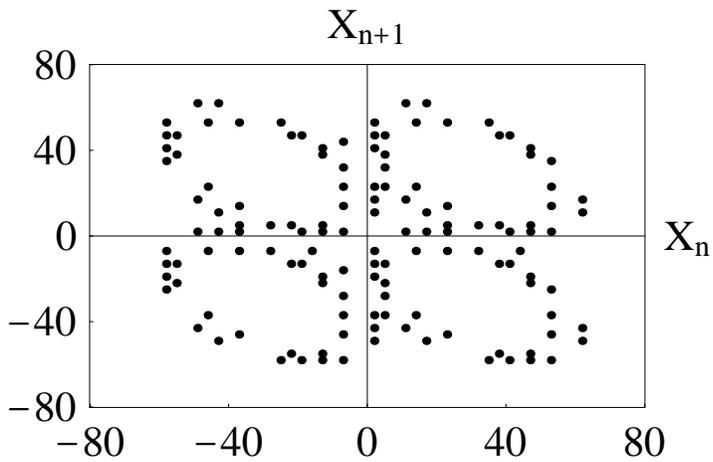}
\end{center}
  \caption{An orbit of (\ref{ultra cutoff}) for $\sigma=3$, $H=10$, $X_0=11$, $X_1=17$. Period is 132.}
  \label{fig4}
\end{figure}

\begin{figure}
\begin{center}
  \epsfile{file=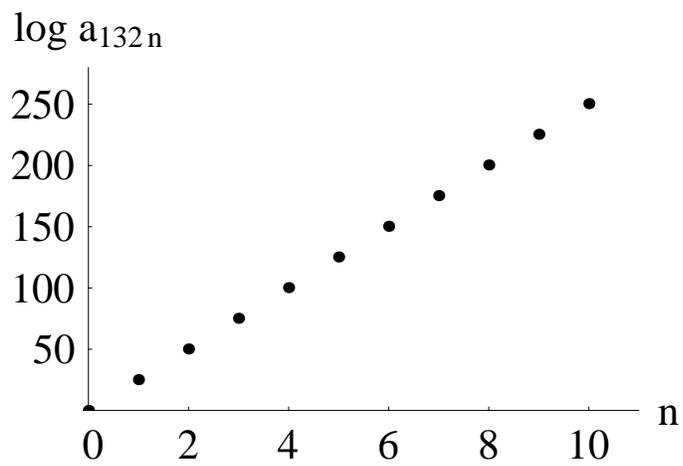}
\end{center}
  \caption{A logarithmic plot of $a_{132n}$ to (\ref{ultra cutoff}) for $\sigma=3$, $H=10$, $X_0=11$, $X'_0=11+\rho$, $X_1=X'_1=17$.  Since slope is about 25, $a_{132n}\sim e^{25n}$.}
  \label{fig5}
\end{figure}

On the other hand, in the case of $\sigma=3$, the growth of perturbation
changes drastically.  
Choosing $H=5$, $X_0=0$ and $X_1=1$, $X_n$ becomes periodic with
period 18 and $a_{18n}=1$, $901$, $326101$, $118047661$,
$42732927181$, $\cdots$.  Clearly $a_n$ shows an exponential growth
and does not grow linearly on $n$.  If we take $H=10$, $X_0=11$ and
$X_1=17$, period is now 132 and $a_{132n}=1$, $84020277977$,
$6297688959151021350515$, $472039455011914887138396062962253$,
$\cdots$.  Fig.~\ref{fig4} shows a figure of the solution orbit and
Fig.~\ref{fig5} shows a relation between $n$ and $\log a_{132n}$, both
in the latter case.  From Fig.~\ref{fig4}, we can see that the orbit
is more complicated than that in $\sigma=2$ case.  From Fig.~\ref{fig5},
$a_{132n}\sim e^{25n}$ and $\lambda\sim25/132$.  Therefore, we can
consider that solutions to equation (\ref{ultra cutoff}) with $\sigma=3$ become
chaotic and thus equation (\ref{ultra cutoff}) is not integrable.
\section{Concluding remarks}
  In this letter, we obtained the following results.
\begin{itemize}
\item[(a)]
  If parameters of ultradiscrete equation are all 0, that is, if
corresponding parameters of difference equation are not
singular, we can consistently transform singularity pattern of difference
equation into ultradiscrete one.  Finite value and singularity of
solution to difference equation correspond to perturbation and finite
value of ultradiscrete solution respectively.
From other point of view, the SC analysis on difference equation can be
regarded as perturbation analysis of solution on ultradiscrete equation.
\item[(b)]
  If parameters of ultradiscrete equation are not 0, that is, if
corresponding parameters of difference equation are singular,
we cannot consistently transform singularity pattern of difference equation
into ultradiscrete one. However, we can observe a growth rate of perturbation,
that is, Lyapunov exponent and check integrability by the exponent.
\end{itemize}

Moreover, we point out an advantage of the above test.  In SC test on
difference equation, we need asymptotic evaluation of a solution
including small parameters.  It often needs a large amount of symbolic
manipulation and becomes hard even with help of a computer.  If we can
transform such a difference equation into ultradiscrete one, we
easily calculate a solution pattern.  Therefore, when we test
integrability of difference equation, it is much easier to test a
corresponding ultradiscrete equation instead.

Finally, we give future problems.
\begin{itemize}
\item[(\romannumeral1)]
  Since integrability test described in (b) is generic, we can use it to
any ultradiscrete equation.  It is easy to conclude a given equation is
chaotic if Lyapunov exponent of a solution from a particular initial
data is positive.  However, it is difficult to judge a given equation is
integrable.  Because we must check Lyapunov exponent is not positive for
ANY initial data.  To avoid this, we need a rule to select initial data to
make the test finite.
\item[(\romannumeral2)]
As mentioned in the introduction, we get some cellular automata using
ultradiscretization when we can restrict values of dependent variables
in a range of integers.~\cite{ultra Burgers}  We can test integrability
of the cellular automata using perturbation.  Therefore, intermediate
values as well as discrete values are significant.  However, cellular
automata are normally defined by binary operations irrelevant to
difference equations and we cannot obtain information on intermediate
values.  Thus, further considerations are required to study cellular
automata using our approach.
\item[(\romannumeral3)]
It is remarkable that Hietarinta and Viallet reported that there are some
difference equations that passes the SC test but not
integrable.~\cite{Hietarin}
Therefore, it is not necessarily true that given equation is integrable
even if it passes the SC test.  (Note that there is no example of integrable
difference equation that does not pass the SC test.) They have proposed to
use the criterion of algebraic entropy instead. Developing the notion of
algebraic entropy for ultradiscrete equations, together with the development
of symbolic manipulation package to deal with max-plus algebra, may be quite
interesting and important.
\end{itemize}
\section*{Acknowledgement} 
The authors are grateful to Prof.~Ryogo Hirota and Mr.~Kinji
Kimura for fruitful discussions and helpful comments.  This work is
partially supported by Grant-in-Aid from the Ministry of Education,
Science and Culture (No.~09750087 and No.~09740164).  \clearpage

\clearpage
\vskip3cm
\clearpage
\vskip3cm
\clearpage
\end{document}